\documentclass{aa}
\usepackage{graphics}

\begin{document}

   \thesaurus{02     % A&A Section 6: Form. struct. and evolut. of stars
              (02.02.1;  % Black hole physics,
               13.07.3;  % Gamma rays: theory,
               13.07.1;  % Gamma rays: bursts,
               13.07.2)}  % Gamma rays: observations.
   \title{On the pair-electromagnetic pulse from an electromagnetic Black Hole surrounded by a Baryonic Remnant}

%%   \subtitle{I. Overviewing the $\kappa$-mechanism}

   \author{Remo Ruffini
          \inst{1}
          \and
          Jay D. Salmonson\inst{2}
	    \and
	    James R. Wilson\inst{2}
	    \and          
	    She-Sheng Xue\inst{1}
%\fnmsep\thanks{Just to show the usage
%          of the elements in the author field}
          }

   \offprints{R. Ruffini (E-Mail: ruffini@icra.it)}
   \institute{I.C.R.A.-International Center for Relativistic Astrophysics
and
Physics Department, University of Rome ``La Sapienza", I-00185 Rome,
Italy
%             email: ruffini@icra.it
	\and Lawrence Livemore National Laboratory, University of
     California, Livermore, California, U.S.A.
%             \thanks{The university of heaven temporarily does not
%                     accept e-mails}
             }

   \titlerunning{On the pair-electromagnetic pulse...}

%  \date{Received 16 May 1998/Accepted 12 July 1998}

   \maketitle

   \begin{abstract}

The interaction of an expanding Pair-Electromagnetic pulse (PEM
pulse) with a shell of baryonic matter surrounding a Black Hole with
electromagnetic structure (EMBH) is analyzed for selected values of
the baryonic mass at selected distances well outside the dyadosphere
of an EMBH. The dyadosphere, the region in which a super critical
field exists for the creation of $e^+e^-$ pairs, is here considered in
the special case of a Reissner-Nordstrom geometry. The interaction of
the PEM pulse with the baryonic matter is described using a simplified
model of a slab of constant thickness in the laboratory frame
(constant-thickness approximation) as well as performing the integration of the general relativistic hydrodynamical
equations. The validation of the constant-thickness approximation, already
presented in a previous paper Ruffini, et al.(1999) for a PEM
pulse in vacuum, is here generalized to the presence of baryonic
matter. It is found that for a baryonic shell of mass-energy less than
1\% of the total energy of the dyadosphere, the
constant-thickness approximation is in excellent agreement with full
general relativistic computations. The approximation breaks down for
larger values of the baryonic shell mass, however such cases are of
less interest for observed Gamma Ray Bursts (GRBs). On the basis of
numerical computations of the slab model for PEM pulses, we describe
(i) the properties of relativistic evolution of a PEM pulse colliding
with a baryonic shell; (ii) the details of the expected emission
energy and observed temperature of the associated GRBs for a given
value of the EMBH mass; $10^3M_\odot$, and for baryonic mass-energies in
the range $10^{-8}$ to $10^{-2}$ the total energy of the dyadosphere.

   \end{abstract}

%
%________________________________________________________________

\section{\it Introduction }\label{introduction}

 That vacuum polarization process {\it \`a la}
Heisenberg-Euler-Schwinger (\cite{he}, \cite{s}) can occur in the field
of a Kerr Newmann EMBH and that they naturally lead to a model for
gamma-ray bursts was pointed out in \cite{dr}. The
basic energetics of the process, governed by the
Christodoulou-Ruffini mass formula, for an EMBH gives as shown in \cite{rc},
\begin{eqnarray}
E^2&=&M^2c^4=\left(M_{\rm ir}c^2 + {Q^2\over2\rho_+}\right)^2+{L^2c^2\over \rho_+^2},\label{em}\\
S&=& 4\pi \rho_+^2=4\pi(r_+^2+{L^2\over c^2M^2})=16\pi\left({G^2\over c^4}\right) M^2_{\rm ir},
\label{sa}
\end{eqnarray}
with
\begin{equation}
{1\over \rho_+^4}\left({G^2\over c^8}\right)\left( Q^4+4L^2c^2\right)\leq 1,
\label{s1}
\end{equation}
where $M, L$ and $Q$ are respectively the mass energy, the angular momentum and the charge of the EMBH and $M_{\rm ir}$ is the irreducible mass, $r_{+}$ is the horizon radius, $\rho_+$ is the quasi-spheroidal cylindrical coordinate of the horizon evaluated at the equatorial plane and $S$ is the horizon surface area. Extreme black holes satisfy the equality in Eq.(\ref{s1}). 
The vacuum polarization process being ``reversible'' transformations in the sense of \cite{rc}
can extract an energy up to 29\% of the mass-energy of an extremal rotating black hole and up to 50\% of the mass-energy of an extremely magnetized and charged black hole.

Although in general such a process is endowed with axial symmetry, in order to clarify the pure interplay of the gravitational and electrodynamical phenomena and also for simplicity, we have examined (\cite{rr} and \cite{prxa}) the limiting cases of such a process in the field of a Reissner-Nordstrom geometry whose spherically symmetric metric is given by
\begin{equation}
d^2s=g_{tt}(r)d^2t+g_{rr}(r)d^2r+r^2d^2\theta +r^2\sin^2\theta
d^2\phi ~,
\label{s}
\end{equation}
where $g_{tt}(r)= - \left[1-{2GM\over c^2r}+{Q^2G\over
c^4r^2}\right] \equiv - \alpha^2(r)$ and $g_{rr}(r)= \alpha^{-2}(r)$. 
The dyadosphere, defined (see Fig. 1 of \cite{prxa}) as the region where the electric field exceeds the
critical value, ${\cal E}_{\rm
c}={m^2c^3\over\hbar e}$ (\cite{he}, \cite{s}), where $m$ and $e$ are the mass
and charge of the electron, extends between the horizon radius
\begin{eqnarray}
r_{+}&=&1.47 \cdot 10^5\mu (1+\sqrt{1-\xi^2})\hskip0.1cm {\rm cm}.
\label{r+}
\end{eqnarray}
where we have introduced the dimensionless mass
and charge parameters $\mu={M\over M_{\odot}}$, $\xi={Q\over (M\sqrt{G})}\le 1$, and the outer radius  
\begin{eqnarray}
r_{\rm ds}
&=&1.12\cdot 10^8\sqrt{\mu\xi} \hskip0.1cm {\rm cm}.
\label{rc}
\end{eqnarray}
Consequently the local number density of
electron and positron pairs created in the dyadosphere as a function of radius
\begin{equation}
n_{e^+e^-}(r) = {Q\over 4\pi r^2\left({\hbar\over
mc}\right)e}\left[1-\left({r\over r_{\rm ds}}\right)^2\right] ~,
\label{nd}
\end{equation}
and their energy density is given by
\begin{equation}
\epsilon(r) = {Q^2 \over 8 \pi r^4} \biggl(1 - \biggl({r \over
r_{\rm ds}}\biggr)^4\biggr) ~, \label{jayet}
\end{equation}
(see Figs.[2] \& [3] of \cite{prxa}). 
The total energy of pairs converted from the static electric energy
and deposited within the dyadosphere is then
\begin{equation}
E_{\rm dya}={1\over2}{Q^2\over r_+}(1-{r_+\over r_{\rm ds}})\left[1-
\left({r_+\over r_{\rm ds}}\right)^2\right] ~,
\label{tee}
\end{equation}
and the total number of electron and position pairs in the dyadosphere is
\begin{equation}
N^\circ_{e^+e^-}\simeq {Q-Q_c\over e}\left[1+{
(r_{\rm ds}-r_+)\over {\hbar\over mc}}\right],
\label{tn}
\end{equation}
where $Q_c={\cal E}_{\rm c}r_+^2$ (see \cite{prxa}).
In the limit of ${r_+\over r_{\rm ds}}\rightarrow 0$, Eq.(\ref{tee}) 
leads to $E_{\rm dya}\rightarrow {1\over2}{Q^2\over r_+}$, which
coincides with the energy 
extractable from EMBHs by reversible processes ($M_{\rm ir}={\rm const.}$), 
namely $E-M_{\rm ir}={1\over2}{Q^2\over r_+}$(see Fig. 4 of \cite{prxa}).

Due to the very large pair density given by Eq.(\ref{nd}) and to
the sizes of the cross-sections for the process $e^+e^-\leftrightarrow
\gamma+\gamma$, the system is expected to thermalize to a plasma
configuration for which
\begin{equation}
N_{e^+}=N_{e^-} \sim N_{\gamma} \sim N^\circ_{e^+e^-},
\label{plasma}
\end{equation}
where $N^\circ_{e^+e^-}$ is the number of $e^+e^-$-pairs created in the 
dyadosphere(see \cite{prxa}). 
These are the initial conditions for the evolution of the dyadosphere. In a previous paper (\cite{rswx99}) we presented the temporal evolution of the dyadosphere in vacuum giving origin to an extremely sharply defined and extremely relativistic expanding pulse of pair and electromagnetical radiation of a constant length in the laboratory frame: the PEM pulse. In this paper we present results of the collision of the PEM pulse with a remnant of baryonic matter surrounding the just formed black hole. We assume the PEM pulse to collide with a shell of baryonic matter of constant density and at a radius of the order of 100 times the radius of the dyadosphere $r_{\rm ds}$ Eq.(\ref{rc}). The shells have this thickness of the order of 10 times $r_{\rm ds}$. The mass-energies of the shells are taken to be $10^{-8}-10^{-2}$ of the total energy of the dyadosphere.

The work contains the following sections: Sect. \ref{hydrodynamic} presents discussions of the hydrodynamical equations of a PEM pulse interacting with the baryonic shell; Sect. \ref{baryon} defines the parameters of the baryonic shells, the behaviour of the PEM pulse colliding with a baryon shell as well as before and after the collision are presented; in Sect. \ref{comparison} the results of a numerical calculation solving the hydrodynamical equations of Sect. \ref{hydrodynamic} are compared to the results of the analytical model of slab approximation for selected parameters.

\section{\it General relativistic hydrodynamical equations for PEM pulse in presence of baryonic matter}\label{hydrodynamic}

The evolution of the PEM pulse in vacuum was treated in a previous work (\cite{rswx99}). We here generalize that treatment to the case in which baryonic matter is present and we outline the relevant relativistic hydrodynamic equations. As in the previous treatment (\cite{rswx99}), we assume the plasma fluid of $e^+e^-$-pairs, photons and baryonic matter to be described by the stress-energy tensor
\begin{equation}
T^{\mu\nu}=pg^{\mu\nu}+(p+\rho)U^\mu U^\nu
\label{tensor}
\end{equation}
where $\rho$ and $p$ are respectively the total proper energy density
and pressure in the comoving frame of the plasma fluid. The $U^\mu$ is the four-velocity
of the plasma fluid.  We have
\begin{equation}
g_{tt}(U^t)^2+g_{rr}(U^r)^2=-1 ~,
\label{tt}
\end{equation}
where $U^r$ and $U^t$ are the radial and temporal contravariant
components of the 4-velocity.  The conservation law of baryon number
can be expressed as a function of the proper baryon number density
$n_B$
\begin{eqnarray}
(n_B U^\mu)_{;\mu}&=& g^{-{1\over2}}(g^{1\over2}n_B
U^\nu)_{,\nu}\nonumber\\
&=&(n_BU^t)_{,t}+{1\over r^2}(r^2 n_BU^r)_{,r}=0 ~.
\label{contin}
\end{eqnarray}
The radial component of the energy-momentum conservation law of the plasma fluid reduces to 
\begin{eqnarray}
&&{\partial p\over\partial r}+{\partial \over\partial t}\left((p+\rho)U^t U_r\right)+{1\over r^2} { \partial
\over \partial r}  \left(r^2(p+\rho)U^r U_r\right)\nonumber\\
&&-{1\over2}(p+\rho)\left[{\partial g_{tt}
 \over\partial r}(U^t)^2+{\partial g_{rr}
 \over\partial r}(U^r)^2\right] =0 ~.
\label{cmom2}
\end{eqnarray}
The component of the energy-momentum conservation law of the plasma fluid equation
along a flow line is
\begin{eqnarray}
U_\mu(T^{\mu\nu})_{;\nu}&=&-(\rho U^\nu)_{;\nu}
-p(U^\nu)_{;\nu},\nonumber\\ &=&-g^{-{1\over2}}(g^{1\over2}\rho
U^\nu)_{,\nu} - pg^{-{1\over2}}(g^{1\over2} U^\nu)_{,\nu}\nonumber\\
&=&(\rho U^t)_{,t}+{1\over r^2}(r^2\rho
U^r)_{,r}\nonumber\\
&+&p\left[(U^t)_{,t}+{1\over r^2}(r^2U^r)_{,r}\right]=0 ~.
\label{conse1}
\end{eqnarray}
Defining the total proper internal energy density $\epsilon$ and the baryonic mass density $\rho_B$ in the comoving frame of the plasma fluid,
\begin{equation}
\epsilon \equiv \rho - \rho_B,\hskip0.5cm \rho_B\equiv n_Bmc^2 ~,
\label{cpp}
\end{equation} 
and using the law of baryon-number conservation (\ref{contin}), from
Eq. (\ref{conse1}) we have
\begin{equation}
(\epsilon U^\nu)_{;\nu} +p(U^\nu)_{;\nu}=0 ~.
\label{conse'}
\end{equation}
Recalling that ${dV\over d\tau}=V(U^\mu)_{;\mu}$, where $V$ is the
comoving volume and $\tau$ is the proper time for the plasma fluid, we have
along each flow line
\begin{equation}
{d(V\epsilon)\over d\tau}+p{dV\over d\tau}={dE\over d\tau}+p{dV\over
d\tau}=0 ~,
\label{f'}
\end{equation}
where $E=V\epsilon$ is total proper internal energy of the plasma fluid.
We represent the equation of state by the introduction of a thermal index
$\Gamma(\rho,T)$
\begin{equation}
\Gamma = 1 + { p\over \epsilon} ~.
\label{state}
\end{equation}

We now turn to the analysis of $e^+e^-$ pairs initially created in the dyadosphere and ionized
electrons contained in the baryonic matter. Let $n_{e^-}$ and $n_{e^+}$  be the proper number densities of electrons and positrons associated to pairs created, and $n^b_{e^-}$ the proper number densities of ionized electrons, we clearly have
\begin{equation}
n_{e^-}=n_{e^+}=n_{\rm pair},\hskip0.5cm n^b_{e^-}=\bar Z n_B,
\label{eee}
\end{equation}
where $n_{\rm pair}$ is the number of $e^+e^-$ pairs and $\bar Z$ the average atomic number ${1\over2}<\bar Z< 1$ ($\bar Z=1$ for hydrogen atom and $\bar Z={1\over2}$ for general baryonic matter). The rate equation for electrons and positrons gives,
\begin{eqnarray}
(n_{e^+}U^\mu)_{;\mu}&=&(n_{e^+}U^t)_{,t}+{1\over r^2}(r^2 n_{e^+}U^r)_{,r}\nonumber\\
&=&\overline{\sigma v} \big[(n_{e^-}(T)+n^b_{e^-}(T))n_{e^+}(T)\nonumber\\
& - &(n_{e^-}+n^b_{e^-})n_{e^+}\big],
\label{e+contin}\\
(n_{e^-}U^\mu)_{;\mu}&=&(n_{e^-}U^t)_{,t}+{1\over r^2}(r^2 n_{e^-}U^r)_{,r}\nonumber\\
&=&\overline{\sigma v} \left[n_{e^-}(T)n_{e^+}(T) - n_{e^-}n_{e^+}\right],
\label{e-contin}\\
(n^b_{e^-}U^\mu)_{;\mu}&=&(n^b_{e^-}U^t)_{,t}+{1\over r^2}(r^2 n^b_{e^-}U^r)_{,r}\nonumber\\
&=&\overline{\sigma v} \left[n^b_{e^-}(T)n_{e^+}(T) - n^b_{e^-}n_{e^+}\right],
\label{tbe-contin}
\end{eqnarray}
where $\overline{\sigma v}$ is the mean of the product of the annihilation cross-section and the thermal velocity of the electrons and positrons,
 $n_{e^\pm}(T)$ are the proper number densities of electrons and positrons associated to the pairs, given
by appropriate Fermi integrals with zero chemical potential, and $n^b_{e^-}(T)$ is the proper number density of ionized electrons, given
by appropriate Fermi integrals with non-zero chemical potential $\mu_e$, at an appropriate equilibrium temperature $T$. These rate equations can be reduced to 
\begin{eqnarray}
(n_{e^\pm}U^\mu)_{;\mu}&=&(n_{e^\pm}U^t)_{,t}+{1\over r^2}(r^2 n_{e^\pm}U^r)_{,r}\nonumber\\
&=&\overline{\sigma v} \big[n_{e^-}(T)n_{e^+}(T)- n_{e^-}n_{e^+}\big],
\label{econtin}\\
(n^b_{e^-}U^\mu)_{;\mu}&=&(n^b_{e^-}U^t)_{,t}+{1\over r^2}(r^2 n^b_{e^-}U^r)_{,r}=0,
\label{becontin}\\
Frac&\equiv&{n_{e^\pm}\over n_{e^\pm}(T)}={n^b_{e^-}(T)\over n^b_{e^-}}.
\label{be-contin}
\end{eqnarray}
Equation (\ref{becontin}) is just the baryon-number conservation law (\ref{contin}) and (\ref{be-contin}) is a relationship obeyed $n_{e^\pm}, n_{e^\pm}(T)$ and $n^b_{e^-}, n^b_{e^-}(T)$.
 
The equilibrium temperature $T$ is determined by the thermalization processes occurring in the expanding plasma fluid with a total proper energy density $\rho$, governed by the hydrodynamical equations (\ref{contin},\ref{cmom2},\ref{conse1}). We have
\begin{equation}
\rho = \rho_\gamma + \rho_{e^+}+\rho_{e^-}+\rho^b_{e^-}+\rho_B,
\label{eeq}
\end{equation}
where $\rho_\gamma$ is the photon energy density, $\rho_B\simeq m_Bc^2n_B$ is the baryonic mass density which is considered to be non relativistic in the range of temperature $T$ under consideration, and $\rho_{e^\pm}$ is the proper energy density of electrons and positrons pairs given by,
\begin{equation}
\rho_{e^\pm}= {n_{e^\pm}\over n_{e^\pm}(T)}\rho_{e^\pm}(T),
\label{hat}
\end{equation}
where $n_{e^\pm}$ is obtained by integration of Eq.(\ref{econtin}) and $\rho_{e^\pm}(T)$ is the proper energy density of electrons(positrons) obtained from zero chemical potential Fermi integrals at the equilibrium temperature $T$. Whereas, $\rho^b_{e^-}$ is the energy density of the ionized electrons, obtained by the ionization of baryonic matter, 
\begin{equation}
\rho^b_{e^-}= {n^b_{e^-}\over n^b_{e^-}(T)}\rho^b_{e^-}(T),
\label{bhat}
\end{equation}
where $n^b_{e^-}$ is obtained by integration of Eq.(\ref{becontin}) and $\rho_{e^-}(T)$ is the proper energy density of ionized electrons obtained from an appropriate Fermi integral of non-zero chemical potential $\mu_e$ at the equilibrium temperature $T$.

Having intrinsically defined the equilibrium temperature $T$ in Eq.(\ref{eeq}), we can also, analogously, evaluate the total pressure 
\begin{equation}
p = p_\gamma + p_{e^+}+p_{e^-}+p^b_{e^-}+p_B,
\label{eep}
\end{equation}
where $p_\gamma$ is the photon pressure, $p_{e^\pm}$ and $p^b_{e^-}$
given by,
\begin{eqnarray}
p_{e^\pm}&=& {n_{e^\pm}\over n_{e^\pm}(T)}p_{e^\pm}(T),
\label{hat'}\\
p^b_{e^-}&=& {n^b_{e^-}\over n^b_{e^-}(T)}p^b_{e^-}(T),
\label{bhat'}
\end{eqnarray}
where the pressures $p_{e^\pm}(T)$ are determined by zero chemical potential Fermi integrals, and $p^b_{e^-}(T)$ is the pressure of the ionized electrons, evaluated by an appropriate Fermi integral of non-zero chemical potential $\mu_e$ at the equilibrium temperature $T$. In Eq.(\ref{eep}), the ions pressure $p_B$ is negligible by comparison with the pressures $p_{\gamma, e^\pm, e^-}(T)$, since baryons and ions are expected to be non-relativistic in the range of temperature $T$ under consideration. Finally, using Eqs.(\ref{eeq},\ref{eep}), we compute the thermal factor $\Gamma$ of the equation of state (\ref{state}). The calculation is continued as the plasma fluid expands,
cools and the $e^+e^-$ pairs recombine, until it becomes optically
thin:
\begin{equation} 
\int_R dr(n_{e^\pm}+\bar
Zn_B)\sigma_T\simeq O(1),
\label{thin}
\end{equation}
where $\sigma_T =0.665\cdot 10^{-24}
{\rm cm^2}$ is the Thomson cross-section and the integration is over the radial interval of the PEM-pulse in the
comoving frame. At this point the energy is virtually entirely in the
form of free-streaming photons and the calculation is stopped. 

\section{\it The quasi-analytical simplified model based on the constant-thickness approximation}\label{baryon}
	
The PEM pulse expansion in the absence of baryonic matter has been discussed in a previous paper (\cite{rswx99}) where the quasi-analytical approach of an expanding shell of constant thickness in the laboratory frame was adopted and validated by comparison with the numerical integration of the general relativistic hydrodynamical equations. We here generalize these results by examining the collision of the PEM pulse with baryonic matter and adopting the constant-thickness approximation both for the description of the collision and the further expansion of the PEM pulse by a simplified approach to the system of equations outlined in the previous paragraph.

We first recall the main results of the PEM pulse expanding in vacuum: 
we indicate by $U(r)=U_r={\rm const.}$ the four-velocity of the slab and by  ${\cal D}=r_{\rm ds}-r_+$  the constant width of the slab in the laboratory frame of the plasma fluid, the average bulk relativistic gamma-factor $\bar\gamma$ is,
\begin{equation}
\bar\gamma=\sqrt{1+U_r^2},\hskip0.5cm V_r={U_r\over\bar\gamma}.
\label{sl1}
\end{equation}
The evolution of the slab is governed by the total energy and entropy conservations, which are cast into the following equations as a function of the coordinate volume 
of the plasma fluid expanding from ${\cal V}_\circ$ to ${\cal V}$, 
\begin{eqnarray}
{\bar\epsilon_\circ\over \bar\epsilon} &=& 
\left({V\over V_\circ}\right)^\Gamma=
\left({ {\cal V}\over  {\cal V}_\circ}\right)^\Gamma\left({\bar\gamma
\over \bar\gamma_\circ}\right)^\Gamma,
\label{scale}\\
\bar\gamma &=&\bar\gamma_\circ\sqrt{{\bar\epsilon_\circ{\cal V}_\circ
\over\bar\epsilon{\cal V}}},
\label{result}\\
{\partial \over \partial t}(N_{e^\pm}) &=& -N_{e^\pm}{1\over{\cal V}}{\partial {\cal V}\over \partial t}+\overline{\sigma v}{1\over\bar\gamma^2}  (N^2_{e^\pm} (T) - N^2_{e^\pm}),
\label{paira}
\end{eqnarray}
where the proper volume $V$ of the plasma fluid
$V=\bar\gamma {\cal V}$ and the thermal index $\Gamma$ Eq.(\ref{state}), a slowly-varying function of the state with values around 4/3, has been approximately assumed to be constant.
The coordinate number density of $e^\pm$-pairs in equilibrium is $ N_{e^\pm}(T)\equiv\bar\gamma n_{e^\pm}(T)$ and the coordinate number density of $e^\pm$-pairs $ N_{e^\pm}\equiv\bar\gamma n_{e^\pm}$. These equations have already been numerically integrated (\cite{rswx99}).

The baryonic matter remnant is assumed to be distributed well outside the dyadosphere in a shell of thickness $\Delta$ between an inner radius $r_{\rm in}$ and an outer radius $r_{\rm out}$ at a distance from the EMBH at which the original PEM pulse expanding in vacuum has not yet reached transparency,
\begin{equation}
r_{\rm in}=100r_{\rm ds},\hskip 0.5cm \Delta = 10r_{\rm ds},\hskip 0.5cm r_{\rm out}=r_{\rm in}+\Delta.
\label{bshell}
\end{equation}
The total baryonic mass ($M_B=N_Bm_pc^2$) is assumed to be a  fraction of the dyadosphere initial total
energy $(E_{\rm dya})$. The total baryon-number ($N_B$) is then given by 
\begin{equation}
N_B=B {E_{\rm dya}\over m_pc^2}.
\label{chimical1}
\end{equation}
where  $B$ is a parameter in the range $10^{-8}-10^{-2}$ and where $m_p$ is the proton mass.
The baryon number density $n^\circ_B$ is assumed to be a constant,
\begin{equation}
\bar n^\circ_B={N_B\over V_B},\hskip0.5cm \bar\rho^\circ_B=m_p\bar n^\circ_B.
\label{bnd}
\end{equation}
 
As the PEM pulse reaches the region $r_{\rm in}<r<r_{\rm out}$, it interacts with the baryonic matter which is assumed to be at rest. In our simplified quasi-analytic model we make the following assumptions to describe this interaction: 

\begin{itemize}

\item
the PEM pulse does not change its geometry during the interaction;

\item
the collision between the PEM pulse and the baryonic matter is assumed to be inelastic,

\item
the baryonic matter reaches thermal equilibrium with the photons and pairs of the PEM pulse.

\end{itemize}

These assumptions are valid if: (i) the total energy of the PEM pulse is much larger than the total mass-energy of baryonic matter $M_B$, $10^{-8}<E_{\rm dya}/M_B <10^{-2}$, and (ii) the comoving number density ratio $n_{e^+e^-}/n^\circ_B$  of pairs and baryons at the moment of collision is extremely high (e.g., $10^6 <n_{e^+e^-}/ n^\circ_B <10^{12}$, and (iii)  the PEM pulse has a large value of Lorentz factor ($600<\bar\gamma < 4000$).  
  
In the collision process between the PEM pulse and the baryonic matter at $r_{\rm out}>r>r_{\rm in}$ , we impose the total energy and momentum conservations. We consider the collision process between two radii $r_2,r_1$, satisfying 
$r_{\rm out}>r_2>r_1>r_{\rm in}$ and $r_2-r_1\ll \Delta$. The amount of baryonic mass acquired by the PEM pulse is
\begin{equation}
\Delta M = {M_B\over V_B}{4\pi\over3}(r_2^3-r_1^3) ,
\label{mcc}
\end{equation}
where $M_B/ V_B$ is the mean-density of baryonic matter at rest.
The conservation of total energy leads to the estimate of the corresponding quantities before (with ``$\circ$") and after such a collision  
\begin{equation}
(\Gamma\bar\epsilon_\circ + \bar\rho^\circ_B)\bar\gamma_\circ^2{\cal V}_\circ + \Delta M = (\Gamma\bar\epsilon + \bar\rho_B + {\Delta M\over V} + \Gamma\Delta\bar\epsilon)\bar\gamma^2{\cal V},
\label{ecc}
\end{equation}
where $\Delta\bar\epsilon$ is the corresponding increase of internal energy due to the collision. Similarly the momentum-conservation gives
\begin{equation}
(\Gamma\bar\epsilon_\circ + \bar\rho^\circ_B)\bar\gamma_\circ U^\circ_r{\cal V}_\circ = (\Gamma\bar\epsilon + \bar\rho_B + {\Delta M\over V} + 
\Gamma\Delta\bar\epsilon)\bar\gamma U_r{\cal V},
\label{pcc}
\end{equation}
where radial component of the four-velocities of the PEM pulse $U^\circ_r=\sqrt{\bar\gamma_\circ^2-1}$ and $\Gamma$ thermal index. 
We then find 
\begin{eqnarray}
\Delta\bar\epsilon & = & {1\over\Gamma}\left[(\Gamma\bar\epsilon_\circ + \bar\rho^\circ_B) {\bar\gamma_\circ U^\circ_r{\cal V}_\circ \over \bar\gamma U_r{\cal V}} - (\Gamma\bar\epsilon + \bar\rho_B + {\Delta M\over V})\right],\label{heat}\\
\bar\gamma & = & {a\over\sqrt{a^2-1}},\hskip0.5cm a\equiv {\bar\gamma_\circ  \over  
U^\circ_r}+ {\Delta M\over (\Gamma\bar\epsilon_\circ + \bar\rho^\circ_B)\bar\gamma_\circ U^\circ_r{\cal V}_\circ}.
\label{dgamma}
\end{eqnarray}
These equations determine the gamma-factor $\bar\gamma$ and the internal energy density $\bar\epsilon=\bar\epsilon_\circ +\Delta\bar\epsilon$ in the capture process of baryonic matter by the PEM pulse.
After collision ($r>r_{\rm out}$), the further adiabatic expansion of PEM pulse is described by the total baryon number, energy and entropy conservations, i.e., the following hydrodynamical equations which generalize those derived in our previous paper (\cite{rswx99}) with $\rho_B\not=0$,
\begin{eqnarray}
{\bar n_B^\circ\over \bar n_B}&=& { V\over  V_\circ}={ {\cal V}\bar\gamma
\over {\cal V}_\circ\bar\gamma_\circ},
\label{be'}\\
{\bar\epsilon_\circ\over \bar\epsilon} &=& 
\left({V\over V_\circ}\right)^\Gamma=
\left({ {\cal V}\over  {\cal V}_\circ}\right)^\Gamma\left({\bar\gamma
\over \bar\gamma_\circ}\right)^\Gamma,
\label{scale1'}\\
\bar\gamma &=&\bar\gamma_\circ\sqrt{{(\Gamma\bar\epsilon_\circ+\rho^\circ_B){\cal V}_\circ
\over(\Gamma\bar\epsilon+\bar\rho_B) {\cal V}}},
\label{result1'}\\
{\partial \over \partial t}(N_{e^\pm}) &=& -N_{e^\pm}{1\over{\cal V}}{\partial {\cal V}\over \partial t}+\overline{\sigma v}{1\over\bar\gamma^2}  (N^2_{e^\pm} (T) - N^2_{e^\pm}).
\label{paira'}
\end{eqnarray}
In these equations ($r>r_{\rm out}$) the comoving baryonic mass- and number densities are $\bar\rho_B=M_B/V$ and $\bar n_B=N_B/V$, where $V$ is the comoving volume of the PEM pulse. The integration is continued untill the transparency condition in Eq.(\ref{thin}) is reached.
%                                                One column figure
%----------------------------------------------------------- fig1
   \begin{figure}
   \resizebox{\hsize}{8cm}{\includegraphics{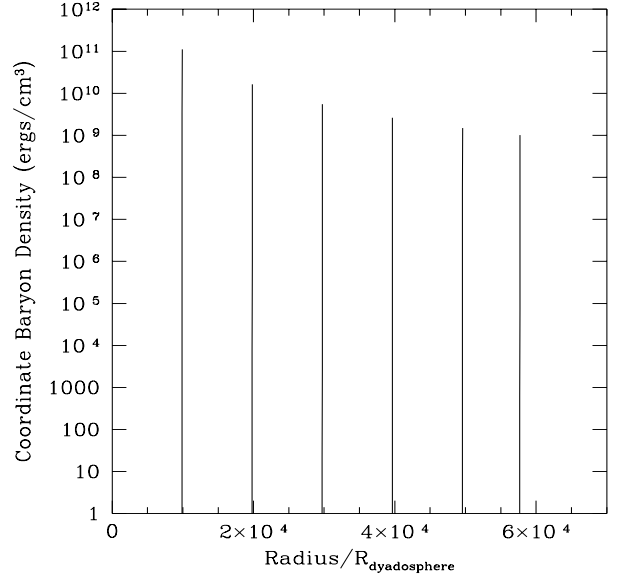}}
      \caption[]{A sequence of snapshots of coordinate baryon energy
density is shown from the one-dimensional (1-D) hydrodynamic
calculations.  This run corresponds to the $M_{\rm BH}=10^3M_\odot, \xi=0.1$ EMBH and $B=1.3\cdot 10^{-4}$ baryonic shell.  The spikes are actually narrow,
unresolved shells (see Fig. \ref{b6detailfig}).
\label{b6fig}}
   \end{figure}
%
%______________________________________________________________

%                                                One column figure
%----------------------------------------------------------- fig2
   \begin{figure}
   \resizebox{\hsize}{8cm}{\includegraphics{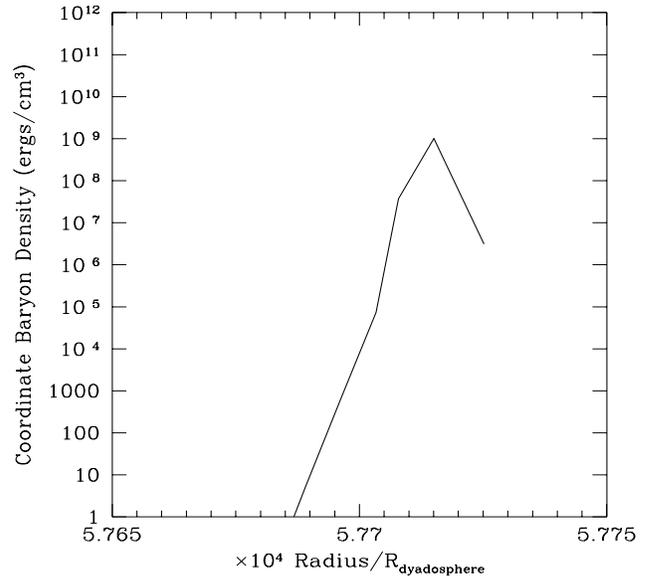}}
      \caption[]{A detail of the last shell of the previous Figure
\ref{b6fig}. In this figure the thickness of the shell is resolved.
\label{b6detailfig}}
   \end{figure}
%
%______________________________________________________________

%                                                One column figure
%----------------------------------------------------------- fig3
   \begin{figure}
   \resizebox{\hsize}{8cm}{\includegraphics{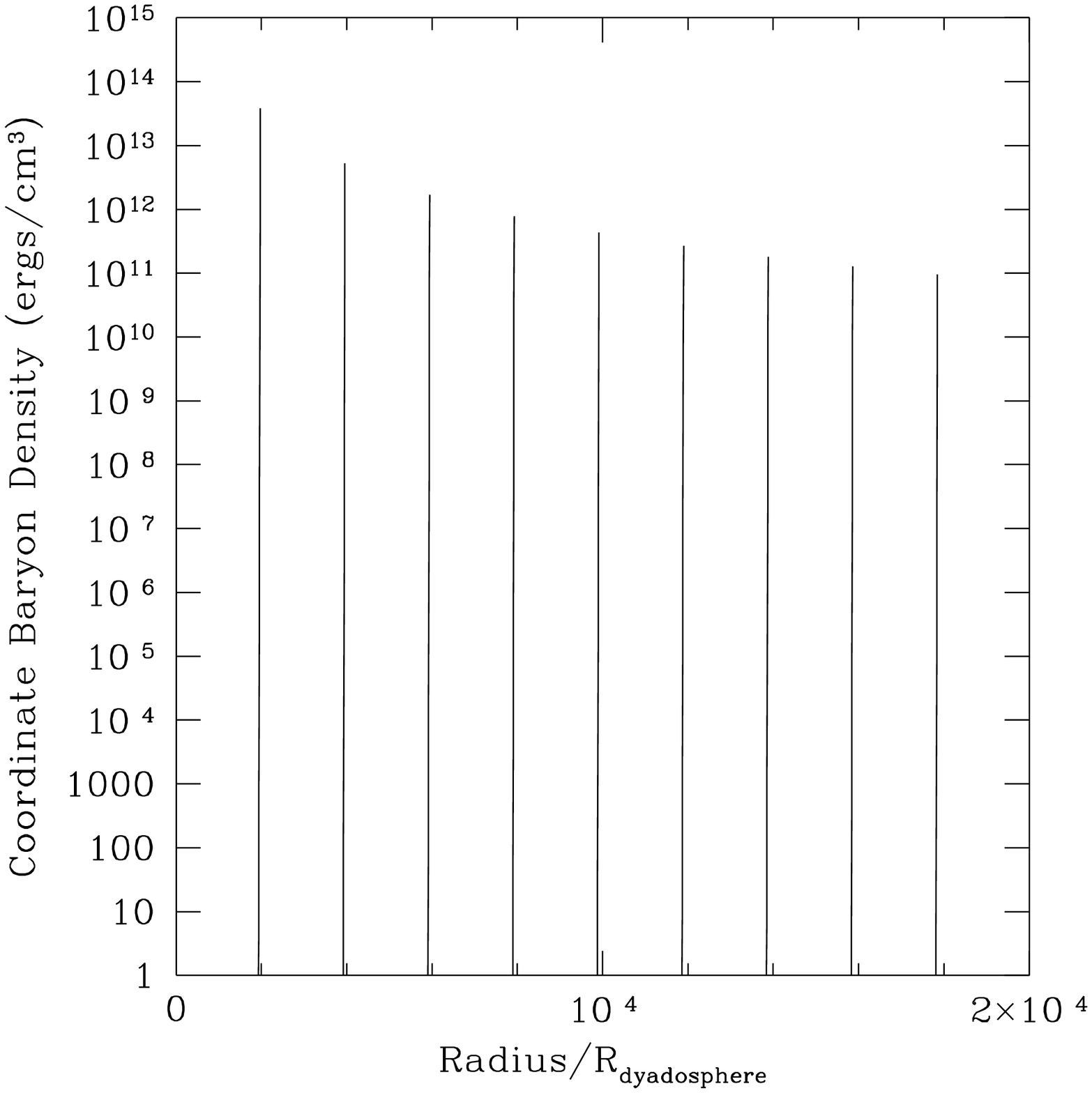}}
      \caption[]{As in Fig. \ref{b6fig} with the $M_{\rm BH}=10^3M_\odot, \xi=0.1$ EMBH and $B=3.8\cdot 10^{-4}$ baryonic shell.
\label{b5.5fig}}
   \end{figure}
%
%______________________________________________________________
%                                                One column figure
%----------------------------------------------------------- fig4
   \begin{figure}
   \resizebox{\hsize}{8cm}{\includegraphics{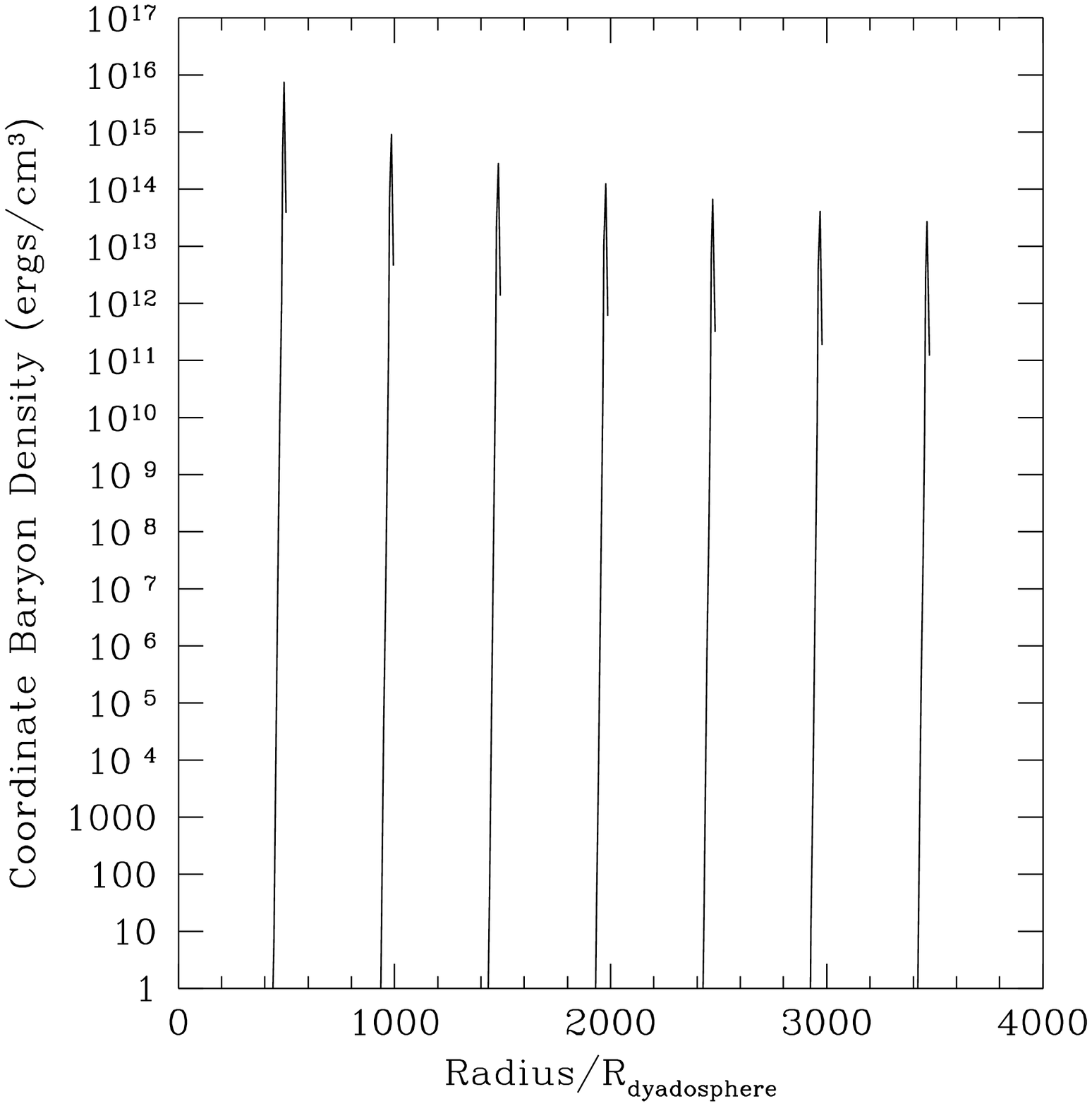}}
      \caption[]{As in Fig. \ref{b6fig} with the $M_{\rm BH}=10^3M_\odot, \xi=0.1$ EMBH and $B=1.3\cdot 10^{-3}$ baryonic shell.
\label{b5fig}}
   \end{figure}
%
%______________________________________________________________
%                                                One column figure
%----------------------------------------------------------- fig5
   \begin{figure}
   \resizebox{\hsize}{8cm}{\includegraphics{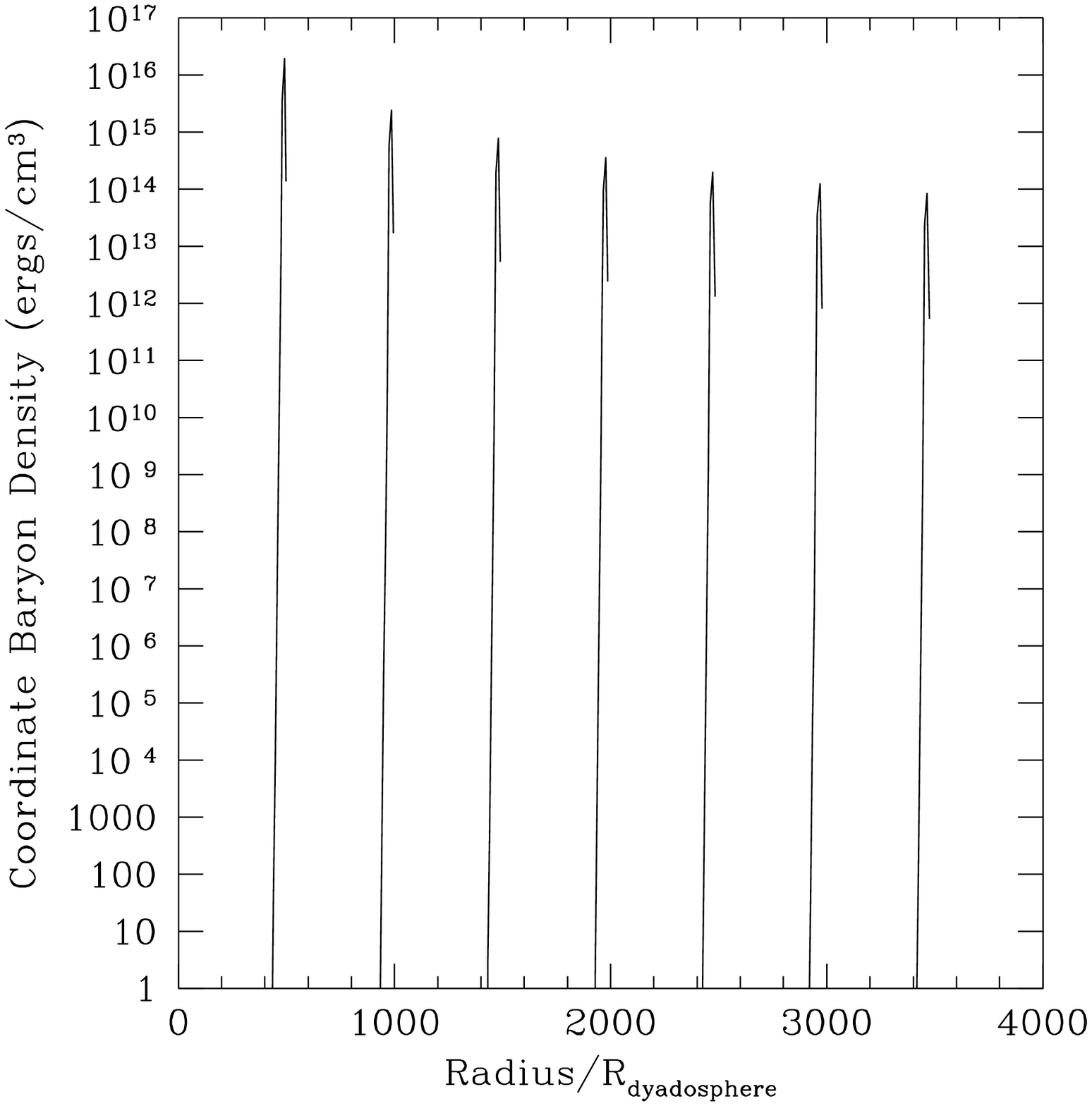}}
      \caption[]{As in Fig. \ref{b6fig} with the $M_{\rm BH}=10^3M_\odot, \xi=0.1$ EMBH and $B=3.8\cdot 10^{-3}$ baryonic shell.
\label{b4.5fig}}
   \end{figure}
%
%______________________________________________________________
%                                                One column figure
%----------------------------------------------------------- fig6
   \begin{figure}
   \resizebox{\hsize}{8cm}{\includegraphics{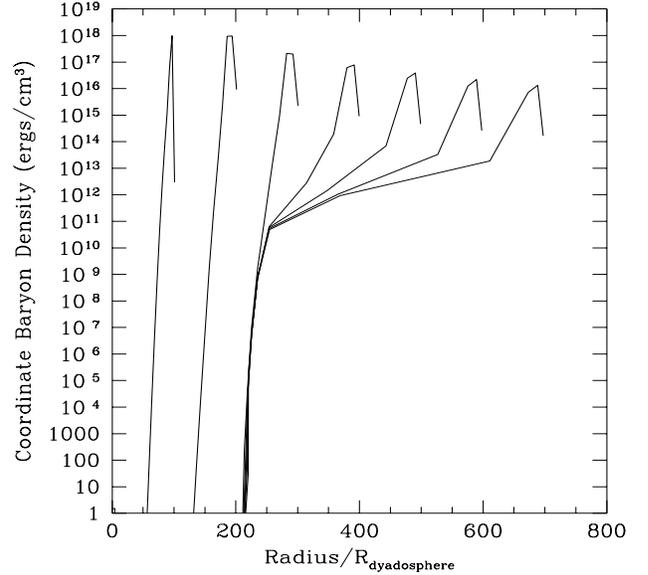}}
      \caption[]{As in Fig. \ref{b6fig} with the $M_{\rm BH}=10^3M_\odot, \xi=0.1$ EMBH and $B=1.3\cdot 10^{-2}$ baryonic shell.  For this baryon shell mass we see
a significant departure from the behavior seen in Figs. \ref{b6fig},
\ref{b5.5fig}, \ref{b5fig}, \ref{b4.5fig}. The baryon shell mass is
significant enough to distort the PEM pulse shell.  Thus the PEM pulse
can no longer be reliably modeled as a shell of constant coordinate
thickness and the comparison of Fig. \ref{twocodecompare} can no
longer be made.
\label{b4fig}}
   \end{figure}
%
%______________________________________________________________

\section{\it Validation of the constant-thickness approximation}\label{comparison}

The numerical integration of the general relativistic equations given in Sect. \ref{hydrodynamic} have already been presented in a series of papers (see \cite{jwg:wmm96}). We can then proceed to compare and contrast the results obtained by the numerical integration of the Eqs.(\ref{contin})-(\ref{conse1}) and the simplified quasi-analytical approach given by Eq.(\ref{scale})-(\ref{paira}), (\ref{ecc})-(\ref{pcc}) and (\ref{be'})-(\ref{paira'}). We select the specific example of an EMBH of $10^3 M_\odot$ with
a charge to mass ratio $\xi=0.1$. From the equations already published (\cite{prxa}), the total energy in the dyadosphere is $3.09\cdot 10^{54}$ergs. The PEM pulse is assumed to collide with a baryonic shell  
of thickness $\Delta =10 r_{\rm ds}$ at rest at a radius
of $100 r_{\rm ds}$. We have considered five different cases: (1) a shell of baryonic mass $2.23 \cdot 10^{-4} M_{\odot}$ ($4 \cdot 10^{50}$ ergs rest energy), from  corresponding to a parameter $B\simeq 1.3\cdot 10^{-4}$; (2) $6.7 \cdot 10^{-4} M_{\odot}$ ($1.2 \cdot 10^{51}$ ergs, $B\simeq 3.8\cdot 10^{-4}$); (3) $2.2 \cdot 10^{-3} M_{\odot}$ ($4 \cdot 10^{51}$ ergs, $B\simeq 1.3\cdot 10^{-3}$); (4) $6.7 \cdot 10^{-3} M_{\odot}$ ($1.2 \cdot 10^{52}$ ergs, $B\simeq 3.8\cdot 10^{-3}$); (5) $2.2 \cdot 10^{-2} M_{\odot}$ ($4 \cdot 10^{52}$ ergs, $B\simeq 1.3\cdot 10^{-3}$).   
The collision between the expanding slab and the baryonic matter shell is treated as an inelastic collision in both calculations. 

\begin{figure}
   \resizebox{\hsize}{8cm}{\includegraphics{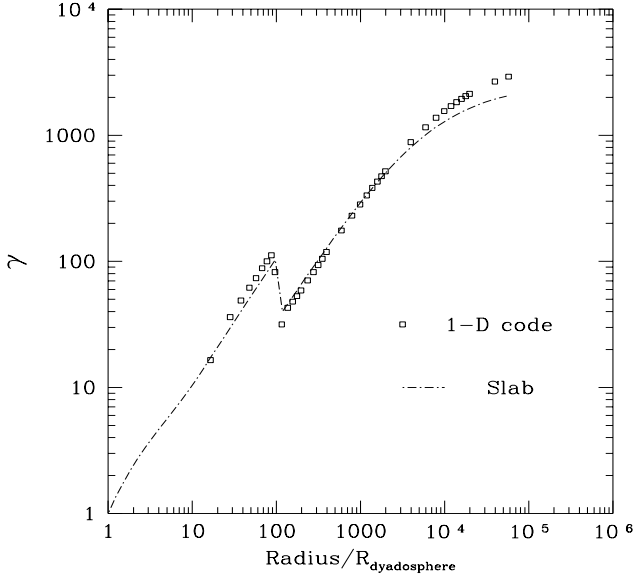}}
      \caption[]{Here we see a comparison of Lorentz factor $\gamma$ for
the one-dimensional (1-D) hydrodynamic calculations and slab
calculations ($M_{\rm BH}=10^3M_\odot, \xi=0.1$ EMBH and $B\simeq 1.3\cdot 10^{-4}$).  
The calculations show good agreement.
\label{twocodecompare}}
   \end{figure}

We first proceed to a qualitative analysis of the evolution of the PEM pulse in Eqs. (\ref{contin})-(\ref{conse1}) from Figs. \ref{b6fig},\ref{b6detailfig} corresponding to $B\simeq 1.3\cdot 10^{-4}$ and Figs. \ref{b5.5fig},\ref{b5fig},\ref{b4.5fig} corresponding to $B\simeq 3.8\cdot 10^{-4}, 1.3\cdot 10^{-3}, 3.8\cdot 10^{-3}$,  we see that the PEM pulse after collision with the baryonic matter shell continues to expand as a one dimensional slab. We see, however, from Figs. \ref{b4fig} corresponding to $B\simeq 1.3\cdot 10^{-2}$ that the expansion after collision becomes much more complex and the constant-thickness approximation ceases to be valid. From this qualitative analysis we now proceed to a quantitative analysis in Fig. \ref{twocodecompare}. 
We compare and contrast the bulk gamma factor as computed from the constant-thickness approximation and the one from the full set of Eqs. (\ref{contin})-(\ref{conse1}). The computation refers to the case $B\simeq 1.3\cdot 10^{-4}$ where an excellent qualitative agreement with the one-dimensional-slab approximation has been found. The extremely good agreement validates the constant-thickness approximation. This agreement has been found up to values of $B$ no larger than $10^{-2}$.
%                                                One column figure
%----------------------------------------------------------- fig7
%
%______________________________________________________________
%                                                One column figure
%----------------------------------------------------------- fig8
   \begin{figure}
   \resizebox{\hsize}{8cm}{\includegraphics{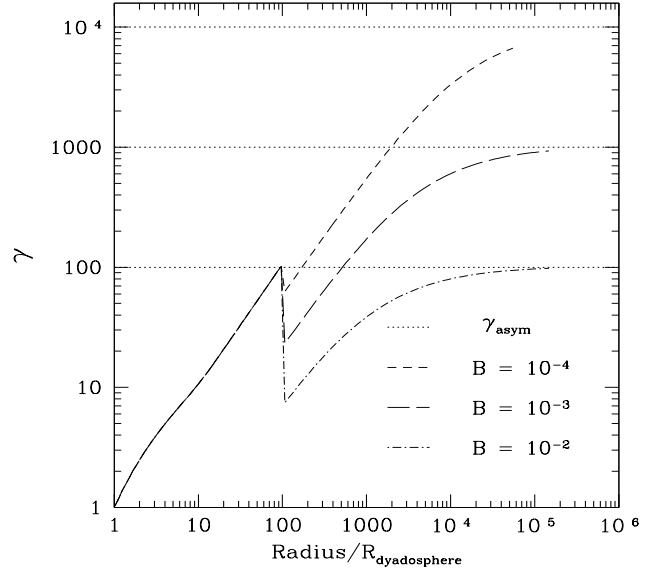}}
      \caption[]{The Lorentz factors are given, as functions of the radius, in units of the dyadosphere radius, for selected values of shell masses given by $B \equiv (M_B c^2)/E_{\rm dya}$ for the typical case $M_{\rm BH}=10^3M_\odot, \xi=0.1$. The asymptotic values $\gamma_{\rm asym} = E_{\rm dya}/ (M_Bc^2)=10^4,10^3,10^2$ are also plotted.
\label{3gamma}}
   \end{figure}
%
%______________________________________________________________

\section{\it Considerations on the GRB structures descending from a constant-thickness approximation }\label{typical}

We now proceed to some specific prediction of GRB features computed by using the constant-thickness approximation and the Eqs. (\ref{scale})-(\ref{paira}), (\ref{ecc})-(\ref{pcc}) and (\ref{be'})-(\ref{paira'}) in the range of validation of this approximation just defined in the previous paragraph.
As an example for clearly showing the evolution of PEM pulses colliding with baryonic matter, we take the following black hole mass and charge, as well as the mass of baryon remnant as a typical case:
\begin{equation}
M_{\rm BH}=10^3M_\odot, \hskip0.5cm \xi=0.1, \hskip0.5cm M_B=10^{-2}E_{\rm dya},
\label{parameter}
\end{equation}
where the total energy of dyadosphere $E_{\rm dya}=3.09\cdot10^{54}$ergs, so the total number of $e^+e^-$-pairs created in the dyadosphere (given by Eq.(\ref{tn})) $N^\circ_{e^+e^-}=1.9\cdot 10^{60}$, and baryonic mass $M_B=1.73\cdot10^{-2}M_\odot$. This baryonic mass is close to the limit of validation of the slab model shown in Sect. \ref{comparison}. We have assumed the baryonic matter at a distance of $r_{\rm in}=100r_{\rm ds}$, very close to the transparency condition of the PEM pulse in vacuum (see \cite{rswx99}).

%                                                One column figure
%----------------------------------------------------------- fig9
   \begin{figure}
   \resizebox{\hsize}{8cm}{\includegraphics{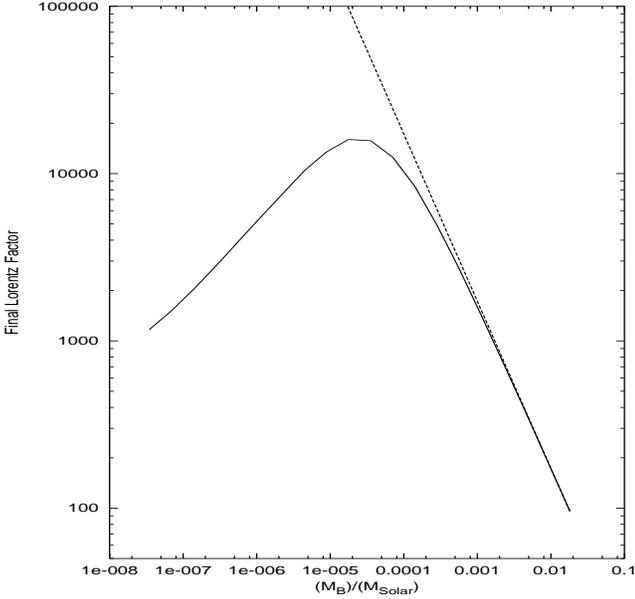}}
      \caption[]{The final Lorentz-factor (the solid line) at the transparent point is plotted as function of baryonic masses $M_B$ in the unit of the solar mass.
The asymptotic value (the dashed line) $E_{\rm dya}/ (M_Bc^2)$ is also plotted.
\label{fgamma}}
   \end{figure}
%
%______________________________________________________________

%                                                One column figure
%----------------------------------------------------------- fig10
   \begin{figure}
   \resizebox{\hsize}{8cm}{\includegraphics{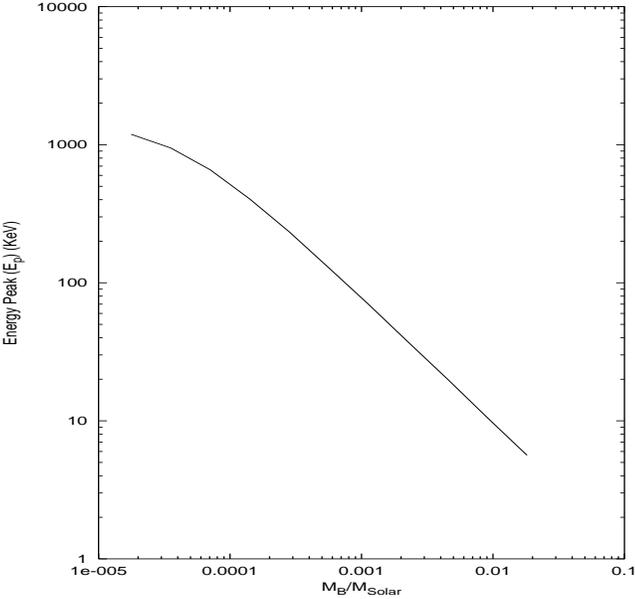}}
      \caption[]{The observed temperature $E_p\sim \bar \gamma T$ in the laboratory frame is plotted as function of baryonic masses $M_Bc^2$ in the unit of the solar mass.
\label{energypeak}}
   \end{figure}
%
%______________________________________________________________

In Fig. \ref{3gamma} we represent the Lorentz Factor of the PEM pulse as a function of the radius for collision with different amounts of baryonic matter, corresponding respectively to $B=10^{-2}$,  
$B=10^{-3}$ and $B=10^{-4}$. The diagram extends to values of the radial coordinate at which the transparency condition given by Eq.(\ref{thin}) is reached. Also represented, for each diagram, is the ``asymptotic'' Lorentz Factor:
\begin{equation}
\bar\gamma_{\rm asym}\equiv {E_{\rm dya}\over M_B c^2}.
\label{asymp}
\end{equation}
The closer the $\bar\gamma$ value approaches, at transparency, the ``asymptotic'' value (\ref{asymp}), the smaller the intensity of the radiation emitted in the burst, and the larger the amount of kinetic energy left in the baryonic matter. This point is further clarified in Fig. \ref{fgamma}, where are plotted the $\bar\gamma$-factor at transparency and the ``asymptotic'' one as functions of the baryonic matter. It is interesting that, for a given EMBH, there is a maximum value of the $\bar\gamma$-factor at transparency. After that maximum value, the energy available for the GRB is smaller in intensity, and at decreasing values of the energy, for increasing values of the baryonic mass.

The temperature in the laboratory frame $\bar\gamma T$ at the transparency point is plotted as a function of the baryonic mass in Fig. \ref{energypeak}: it strongly decreases as the baryonic mass increase. The $\bar\gamma T$ is related to the observed energy-peak of the photon-number spectrum (see e.g., \cite{rswx99}).

%                                                One column figure
%----------------------------------------------------------- fig11
   \begin{figure}
   \resizebox{\hsize}{8cm}{\includegraphics{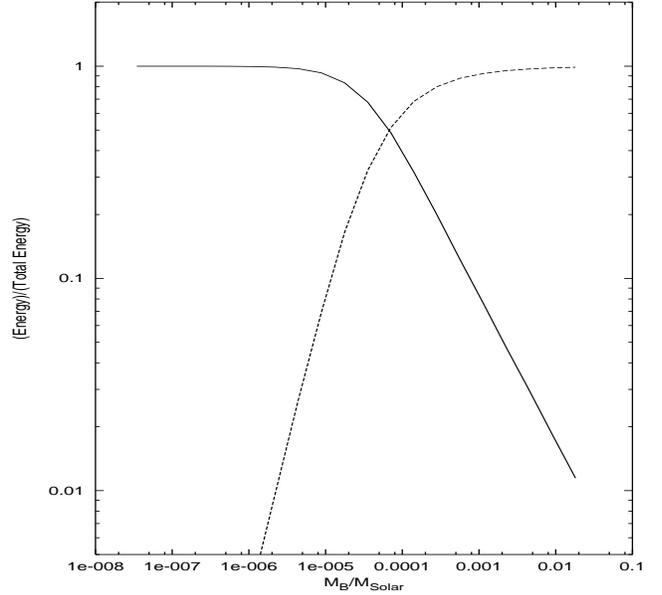}}
      \caption[]{At the transparent point, the energy radiated in the burst (the solid line) and the final kinetic energy (the dashed line) of baryonic matter, in units of the total energy of the dyadosphere, are plotted as functions of baryonic mass $M_Bc^2$ in units of solar mass.
\label{fintkin}}
   \end{figure}
%
%______________________________________________________________
%                                                One column figure
%----------------------------------------------------------- fig12
   \begin{figure}
   \resizebox{\hsize}{8cm}{\includegraphics{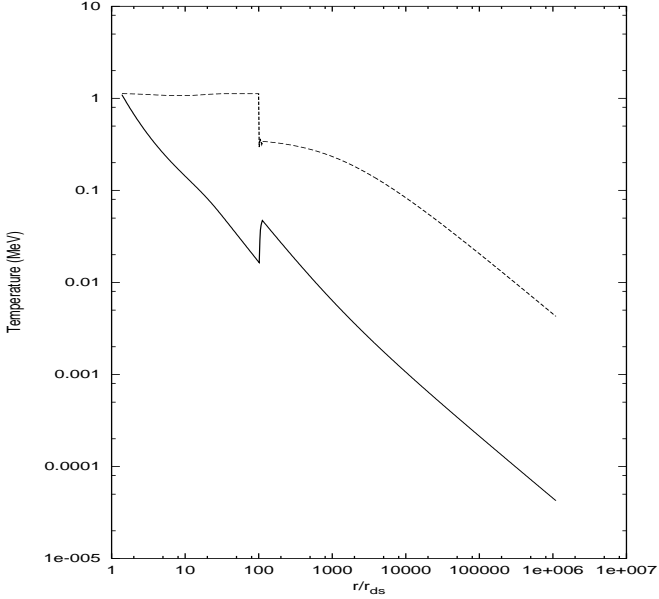}}
      \caption[]{The temperature in the comoving frame $T'$(MeV) (the solid line) and in the laboratory frame $\bar\gamma T'$ (the dashed line) are plotted as functions of the radius in the unit of the dyadosphere radius $r_{\rm ds}$ for the typical case $M_{\rm BH}=10^3M_\odot, \xi=0.1$ and $B=10^{-2}$ (see Sect. \ref{typical} in the text).
\label{tem}}
   \end{figure}
%
%______________________________________________________________
%                                                One column figure
%----------------------------------------------------------- fig13
   \begin{figure}
   \resizebox{\hsize}{8cm}{\includegraphics{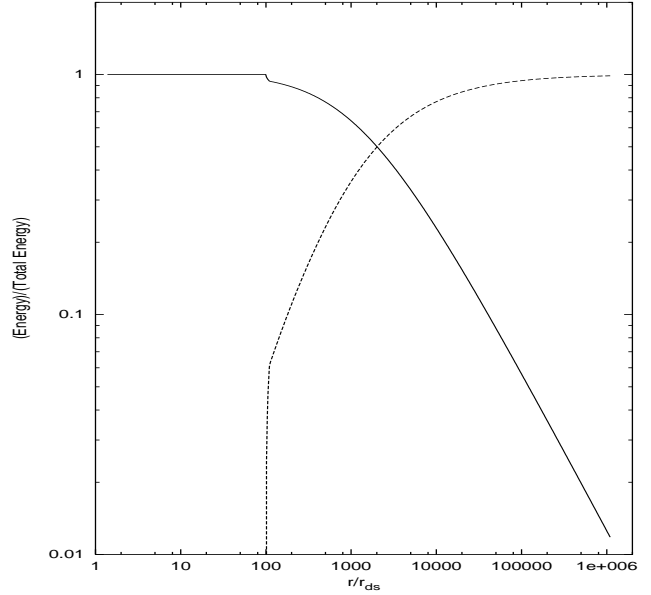}}
      \caption[]{The total energy of the non baryonic components of the PEM pulse (the solid line) and the total kinetic energy of baryonic matter (the dashed line) in the unit of the total energy are plotted as functions of the radius in the unit of the dyadosphere radius $r_{\rm ds}$ for the typical case $M_{\rm BH}=10^3M_\odot, \xi=0.1$ and $B=10^{-2}$.
\label{intkin}}
   \end{figure}
%
%______________________________________________________________
%                                                One column figure
%----------------------------------------------------------- fig14
   \begin{figure}
   \resizebox{\hsize}{8cm}{\includegraphics{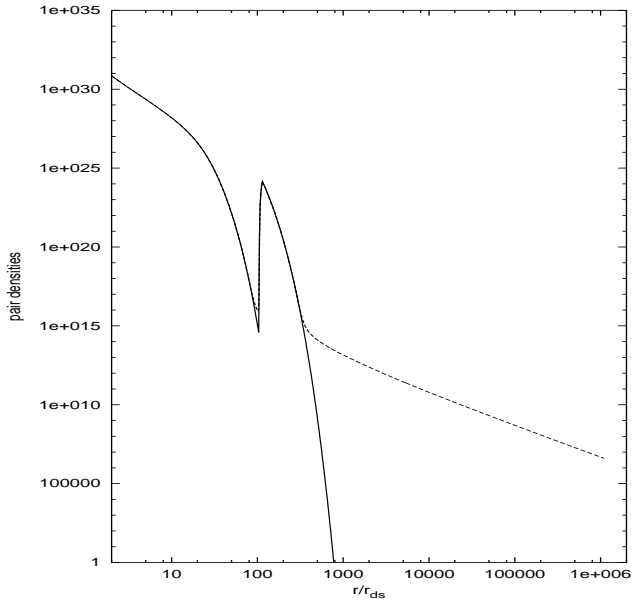}}
      \caption[]{The number densities $n_{e^+e^-}(T)$ (the solid line) computed by the Fermi integral and $n_{e^+e^-}$ (the dashed line) computed by the rate equation (\ref{paira}) are plotted as functions of the radius. $T'\ll m_ec^2$, two curves strongly divergent due to $e^+e^-$-pairs frozen out of the thermal equilibrium. The peak at $r=r_{\rm ds}$ is due to the temperature reheating in the collision.
\label{pair}}
   \end{figure}
%
%______________________________________________________________

%                                                One column figure
%----------------------------------------------------------- fig15
   \begin{figure}
   \resizebox{\hsize}{8cm}{\includegraphics{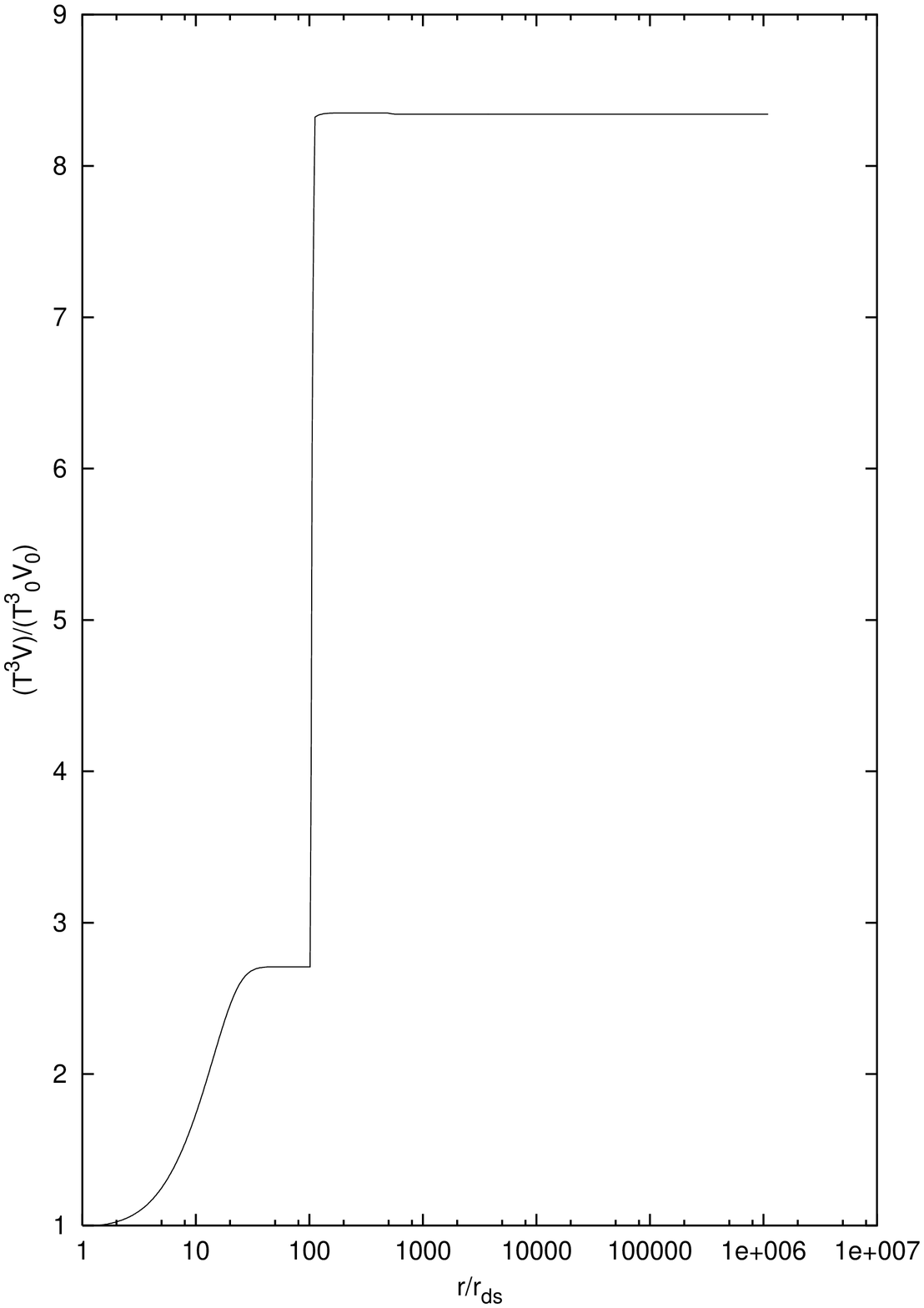}}
      \caption[]{We plot the ratio $(T^3V)/(T^3_\circ V_\circ)$, $T$ being the temperature of the PEM pulse, and $V$ the volume in the comoving frame, as function of the radius in the unit of the dyadosphere radius $r_{\rm ds}$ for the typical case $M_{\rm BH}=10^3M_\odot, \xi=0.1$ and $B=10^{-2}$ (see Sect. \ref{typical} in the text). After the collision, there is a small reheating up to the annihilation of the $e^+e^-$ pairs recreated in the collision (see Fig. \ref{pair}), which cannot be seen on the scale of our plot.
\label{entropy}}
   \end{figure}
%
%______________________________________________________________

We plot in Fig. \ref{fintkin} the energy radiated in the burst and the final kinetic energy of baryonic matter. We find that, for small values of $B$ (around $10^{-8}$), almost all total energy is radiated as GRB (see also our previous paper \cite{rswx99}), and very little energy is left as kinetic energy of baryonic matter as afterglow. While for $B\simeq10^{-2}$ roughly only $10^{-2}$ of the total initial energy of the dyadosphere is radiated away as GRB, and almost all energy is restored as the kinetic energy of the baryonic matter. It is also clear that for $B>10^{-2}$ the intensity of the Burst (see also Fig. \ref{3gamma}) and the observed radiation frequency drifted to smaller and smaller values and are of little astrophysical interest from the point of view of GRBs. For such values, the energy of the dyadosphere is transferred practically totally to the bulk kinetic energy of the baryonic matter. $B=10^{-2}$ is also the limit of the validation of our computations based on the analytical slab model, described in Sect. \ref{baryon}. For values of baryonic matter in between these two extremal cases ($B=10^{-8}$ and $B=10^{-2}$) the analytical slab model covers the whole range of the observed properties of Gamma Ray Bursts.

We turn now to the thermodynamic parameters relevant in the evolution of the PEM pulse. In Fig. \ref{tem} the temperature of PEM pulse, both in the comoving and in the laboratory frame, are given as a function of the radius for a typical case ($M_{\rm BH}=10^3M_\odot, \xi=0.1$ and $B=10^{-2}$). 

In Fig. \ref{intkin}, we plot the total energy of the non baryonic components of the PEM pulse, which includes both thermal and kinetic energy, and the kinetic energy of the baryonic matter as functions of the radius, for the typical case $M_{\rm BH}=10^3M_\odot, \xi=0.1$ and $B=10^{-2}$. The total energy of the non baryonic components of the PEM pulse is equal to $E_{\rm dya}$ before the collision (see \cite{rswx99}) and drops after the collision. While the kinetic energy of baryonic matter
\begin{equation}
E_{k}=\bar\rho_B V(\bar\gamma -1)
\label{kinetic}
\end{equation}
increases as function of radius for $(r>r_{\rm in})$. The sum of both them is equal to the total energy $E_{\rm dya}=3.09\cdot 10^{54}$ergs during the evolution of the PEM pulse. The value of the total energy of the non baryonic components of the PEM pulse at the transparency point, the ending point of the curve in Fig. \ref{intkin}, is the energy released in the burst. We have discussed this energy as function of baryonic masses in Fig. \ref{fintkin}.

Before and after the collision, the condition of entropy conservation applies, and we have:
\begin{equation}\label{EntrPrima}
S_{\rm before}=\frac{V}{T}\left(\rho_{\gamma}+\rho_{e^\pm}+p_{\gamma}+p_{(e^\pm)}\right)
\end{equation}
\begin{eqnarray}\label{EntrDopo}
S_{\rm after}&=&\frac{V}{T}\big(\rho_{\gamma}+\rho_{e^\pm}+\rho^b_{e^-}+\epsilon_B\nonumber\\
&+&p_{\gamma}+p_{e^\pm}+p_B+p^b_{e^-}\big),
\end{eqnarray}
where $\epsilon_B$ is the thermal energy of baryonic matter, and we can neglect, in Eq. (\ref{EntrDopo}), the pressure of the baryonic matter.
A sudden increase of the entropy occurs during the collision both for the addition of the baryonic matter, and for the thermal reheating due to the inelastic collapse of the PEM pulse with the baryonic matter at rest. From the energy and momentum conservations, we obtain for the values of the $\bar\gamma$ and the temperature during the collision: the proper internal
energy, $E_{\rm int}$, of the plasma increases as
\begin{equation}
dE_{\rm int} = (\gamma - 1) dM_B
\label{Eint}
\end{equation}
and the slab is decelerated, in terms of Lorentz factor $\gamma$, as
\begin{equation}
d\gamma = - \frac{\gamma^2 - 1}{M_Bc^2 + E_{\rm int}} dM_B ~,
\label{gammadecel}
\end{equation} 
as baryon mass, $dM_B$, is incrementally gained. 

Before the collision the PEM pulse expands keeping its temperature in the laboratory frame constant, while its temperature in the comoving frame falls (see \cite{rswx99}). During the collision, a heating of the plasma due to the energy and momentum conservation occurs (see also Fig. \ref{pair}, where reheating process leads to an increment of the number density of $e^+e^-$ pairs). As the system expands further, both the comoving temperature and the temperature in the laboratory frame decreases, since the total energy of the $e^+e^-$ pairs and the photons before the collision is constant and equal to $E_{\rm dya}$, while after the collision
\begin{equation}
E_{\rm dya}=E_{\rm Baryons}+E_{e^+e^-}+E_{\rm photons}
\label{Etotal}
\end{equation}
where $E_{\rm Baryons}$ is only the thermal energy of the baryons.

It is also interesting to monitor the change of temperature in the comoving frame before the collision and after the collision (see also Fig. \ref{entropy}). Before the collision, due to the entropy conservation, in the process of $e^+e^-$ annihilation the factor $\frac{T^3V}{T_\circ^3V_\circ}$ (where the subscript ``$_\circ$" means ``initial value") increases to a value near to $\frac{11}{4}$ (see \cite{ws}; \cite{rswx99}) since the collision occurs at $r=100r_{\rm ds}$, near the condition of transparency. The number of $e^+e^-$ pairs has now been reduced drastically. The further jump in the value of the ratio $\frac{T^3V}{T_\circ^3V_\circ}$ is principally due to the energy and momentum conservation during the inelastic collision. After the collision, there is a small reheating due to the annihilation of the $e^+e^-$ pairs created in the collision (see Fig. \ref{pair}), which cannot be seen on the scale of our plot. Then, the ratio remains constant.

\section{\it Conclusions}\label{conclusions}

By the direct comparison with the numerical integration of the complete relativistic hydrodynamical equations, the use of the constant-thickness approximation has been validated for values of the parameter $B\leq10^{-2}$. For $B\geq 10^{-2}$ the amount of energy released at transparency in the burst decreases, and its energy drifts toward low energy values, which are of little interest for the astrophysics of GRB. We conclude that the constant-thickness approximation is valid for all astrophysically relevant situations for the analysis of the GRB at transparency. Based on this validation we have studied the evolution of a PEM pulse in the presence of selected amounts of baryonic matter. We have studied for a typical case of $M_{\rm BH}=10^3M_\odot, \xi=0.1$ EMBH and $B= 10^{-2}$ baryonic shell, the thermal, the bulk Lorentz factor evolution of the PEM pulse as well as the kinetic energy left over in the baryonic matter after the decoupling of matter and radiation and the emission of the GRB. Additional results corresponding to a larger range of masses and charges of the EMBH and the correlations between the peak energy and the duration of GRBs to be expected in our model will be considered in a future paper, together with the results of analyzing the interaction of the kinetic energy left over in the baryonic matter, after the decoupling of matter and radiation, with the interstellar medium.

\section{\it Acknowledgments}
     
Work by JDS and JRW was performed under the auspices of the U.S. Department of Energy by University of California Lawrence Livermore National Laboratory under contract No. W-7405-Eng-48.

\end{document}